# Approaching the Lower Temporal Limit of Laser-Produced Plasma Sources for Table-Top Soft X-ray NEXAFS Measurements

**Danika Nimlos,[a] Alejandro Arellano,[a] and Scott Kevin Cushing*[a]**

[a] Division of Chemistry and Chemical Engineering
California Institute of Technology
Pasadena, California 91125, United States
* E-mail: scushing@caltech.edu

**Abstract:** The increasing popularity of time-resolved X-ray absorption measurements for understanding dynamics in molecular and material systems has led to many advances in table-top sources for pulsed X-rays. We report on a table top laser-produced plasma (LPP) source that can perform soft X-ray (SXR), near edge X-ray absorption fine structure (NEXAFS) measurements using a laser source with 23 ps pulse duration. The spectrometer's key specifications, such as brilliance, resolution, and stability, are characterized against the more commonly used longer-pulse-duration LPP sources. The 23 ps laser produced approximately an order of magnitude weaker SXR flux than the 8 ns laser for a higher power density due to the smaller total energy absorbed by the plasma. The increased repetition rate and an implemented self-referencing scheme still allowed for high-resolution, synchrotron-like NEXAFS measurements of $Si_3N_4$ and $TiO_2$ thin films with 2.5 minute acquisition times. It was observed that degradation of the gas jet nozzle led to long-term instability of the source, which can be remediated using alternative nozzle designs. This work demonstrates the feasibility of achieving higher temporal resolution in future time-resolved X-ray absorption measurements using table-top laser-produced plasma sources.

## Introduction

X-ray absorption spectroscopy (XAS) is an invaluable tool in material and chemical sciences, used for a wide range of applications from photocatalysis to solid-state batteries.[1–4] X-rays excite element-specific core-level transitions, with the near-edge X-ray absorption fine structure (NEXAFS) providing information on oxidation state, coordination, hybridization, and bonding.[5] NEXAFS measurements are predominantly conducted at large-scale user facilities such as synchrotrons or free electron lasers because they require broadband, high-energy sources. While advances in table-top sources for X-ray measurements have made them an increasingly viable alternative, reported setups still face limitations in the tradeoff between energy range, flux, and temporal resolution.

The increasing demand for XAS measurements underscores the need for continued advances in table-top approaches, particularly for time-resolved studies. High harmonic generation (HHG) and laser-produced plasma (LPP) generation are the most common methods explored for table-top X-ray sources. HHG utilizes single-cycle or near-single-cycle pulses to generate coherent X-rays through a three-step process.[6] Although HHG sources can generate sufficient flux in the extreme ultraviolet regime for routine transient absorption measurements,[7–9] they require complex specialty laboratories, and extending their energy range remains challenging. LPP sources generate incoherent soft or hard X-rays by ionization of a gas, liquid, or solid target.[10–14] Gaseous LPPs have been the most studied because they produce minimal debris, utilize the widely available and cost-effective Nd:YAG laser system, and generate broad-band radiation in the soft X-ray (SXR) regime, striking a balance between cost, brilliance, energy range, and feasibility. However, the temporal resolution of gaseous LPP sources has been typically limited to nanoseconds or in some cases hundreds of picoseconds.[15,16] Improving the temporal resolution of LPP sources could enable the measurement of faster dynamics, such as carrier recombination,[17] polaron dynamics,[18] or charge transfer.[19]

This work explores the lower limits of pulse duration in an LPP for use as a tabletop soft X-ray spectrometer. The time resolution of laser-driven SXR emission from LPP sources is constrained by the thermalization time of free electrons in the plasma. The thermalization time is dictated by the electron plasma frequency, $\omega_p$, which describes the oscillations in electron density due to the competing forces of ionized electrons accelerating away from nuclei and the Coulombic attraction with positive nuclei.[20] Using the root mean square thermal electron velocity $v_e$, determined by the electron temperature $T_e$, the thermal expansion time $t_D$ can be related to the electron plasma frequency such that,

$$t_D \cong \frac{\lambda_D}{v_e} = \left(\frac{\varepsilon_0 k_B T_e}{e^2 n_e} \cdot \frac{m_e}{k_B T_e}\right)^{\frac{1}{2}} = \omega_p^{-1}$$

where $\lambda_D$ is the Debye length, which measures the length of a charge's electrostatic effect, $n_e$ is the electron density of the target, $\varepsilon_0$ is the permittivity of free space, and $m_e$ and $e$ are the mass and charge of an electron, respectively.[21] Using average electron densities of LPPs in gaseous targets as $10^{19}$ e$^-$/cm$^3$,[16,22–24] the expansion time is on the order of picoseconds.[25] For effective bremsstrahlung emission of X-rays, and therefore a usable X-ray continuum for measurement, the driving laser pulse must have a duration longer than the thermal expansion time of the electrons.[20]

LPP sources in gaseous targets have not been demonstrated using laser pulses below a 170 ps pulse width. In this study, we report SXR emission from an LPP source generated with 23 ps pulses. We benchmark our source against an 8 ns Nd:YAG, commonly used in LPP setups.[26–28] Although per-pulse-flux of the ps source is nearly an order of magnitude less than that of the ns source, the increased laser repetition rate and a self-referencing scheme still allow high resolution NEXAFS spectra with measurement times of 2.5 minutes. This is demonstrated with $Si_3N_4$ and $TiO_2$ thin films. The findings of this paper are important for pushing table-top SXR spectroscopy into the tens of picosecond regime, which is critical for measuring liquid phase and solid-state dynamics. Furthermore, the use of an Nd:YAG laser and electronic delay lines can extend measurement timescales from the nanosecond to second in future transient measurements. This approach offers a more space and cost-effective alternative to most HHG based setups.

## Experimental Section

**Figure 1** is a photograph and schematic diagram of the table-top SXR spectrometer. The spectrometer is built in three 9" × 9" × 9" cubic vacuum chambers that can be evacuated to base pressures of $10^{-5}$ Torr. The plasma generation is driven by the 1064 nm fundamental beam of an Nd:YAG laser, either an EKSPLA PL2251B-20 (20 Hz repetition rate, 23 ps pulse duration, 80 mJ pulse energy) or a Spectra-Physics Quanta-Ray PRO (10 Hz repetition rate, 8 ns pulse duration, 800 mJ pulse energy). The 1064 nm beam from the Nd:YAG laser is focused with a 150 mm focal length lens onto a gas jet inside the vacuum chamber. Due to the losses by the mirrors, lens, window, and other optical elements in the beam path, the pulse energy of the ns laser at the gas jet is 640 mJ/pulse, giving a power density of $2.47 \times 10^{13}$ W/cm$^2$. The pulse energy of the ps laser is 67 mJ/pulse at the gas jet, with a power density of $8.96 \times 10^{14}$ W/cm$^2$. Focal spot size is estimated by the respective laser specifications and lens focal length. Input and output energies of the ps laser were measured with a thermopile sensor (Newport 919P-050-26), and those of the ns laser were measured with a Scientech 37-2002 power meter with a 380101 (400-1200 nm) detector head.

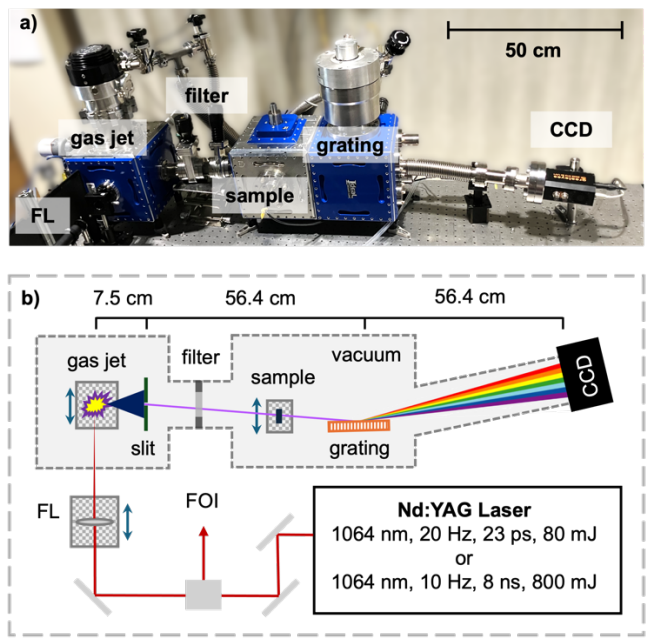

**Figure 1. a)** Instrument photograph and **b)** schematic of the table-top soft X-ray spectrometer with the Nd:YAG laser beam path passing through a faraday optical isolator (FOI), where any backscatter reflected off the plasma is rejected, before passing through a focusing lens (FL) where it is focused onto a pulsed gas jet inside a vacuum chamber. Generated X-rays emitted 90° to the incoming laser are then measured by the SXR spectrometer.

Ar, Kr, or $N_2$ gas with a backing pressure of 175 psi is delivered by a high-speed pulsed solenoid valve (Parker 009-0347-900) using a stainless-steel nozzle with a 0.02" diameter opening. The gas jet is electronically delayed from the main pulse generator in the Nd:YAG laser. The solenoid delivers gas pulses with a duration of $t_{open}$ = 350 μs or 900 μs for the ps and ns lasers, respectively. A filter gate valve (VAT, 01032-CE01-AAV2) separates the plasma generation chamber from the rest of the setup to optimize vacuum pressures and minimize debris from the plasma generation process. During operation of the gas jet, the plasma generation chamber is held at $10^{-3}$ Torr and the rest of the setup remains at $10^{-5}$ Torr.

The X-rays emitted from the plasma pass through either a 50 or 100 μm slit and a 100 nm Al filter enclosed in the filter gate valve, which blocks any out-of-band emission and scattered 1064 nm light from the plasma. Soft X-rays interact with the sample in transmission geometry before being diffracted by an aberration-corrected concave grating (3,600 grooves per mm, 564 mm focal length, Shimadzu, L3600-1-6) and measured by a charged coupled device (CCD, Greateyes, GE 2048 512 BI UV1, 13.5 × 13.5 μm pixel size, 512 × 2048 pixels). For the spectral comparisons of the ns and ps lasers, the CCD was operated at room temperature.

Energy calibration is performed by creating a calibration curve using the Ar IX emission lines at λ = 4.873 nm and 4.918 nm and the N VI emission lines at λ = 2.878 nm, 2.489 nm, 2.377 nm, and 2.302 nm, as documented in the NIST atomic spectra database.[29] The pixel positions for each of these energies were then fit to a quadratic polynomial. After calibration, energy resolution was experimentally measured by fitting the N VI emission line at λ = 2.878 nm to a single-term Gaussian function, such that,

$$f(x) = a_1 \times e^{-\left(x - b_1 / c_1\right)^2}$$

where the coefficient $c_1$ can be used to calculate the FWHM by,
$FWHM = 2\sqrt{\ln 2}\, c_1$.

For the NEXAFS measurements, Kr gas was used due to its quasi-continuous emission spectrum. A 100 μm slit was used to reduce measurement times. A self-referencing scheme, demonstrated previously,[30] utilizes the 4π emission from the plasma to measure SXR transmission of the sample and a reference simultaneously. Self-referencing was employed to reduce the effects of shot-to-shot noise and long-term instabilities of the plasma source. The CCD was cooled to -50 °C to reduce background noise, and dark counts were subtracted from all spectra.

The $TiO_2$ thin film sample was prepared by atomic layer deposition (ALD) on 50 nm thick diamond membranes with 2 mm diameter apertures. Silicon nitride ($Si_3N_4$) membranes were obtained commercially (Norcada Inc., NX10300A).

## Results and Discussion

### Nanosecond vs. Picosecond X-ray Emission

**Figure 2** compares the Ar emission spectra of plasmas generated with 8 ns and 23 ps lasers. Spectra were collected over 1 s integration times (20 and 10 pulses for the ps and ns lasers, respectively) using a 50 μm slit. When integrated over all energies, the ns laser has 4.6 times greater total flux per integration period, which is 9.2 times greater flux per pulse. This difference could be explained by the total absorbed energies of the plasmas for the two laser sources. For the spectra in **Figure 2**, the absorbed energies are 60 mJ and 235 mJ for the ps and ns lasers, respectively.

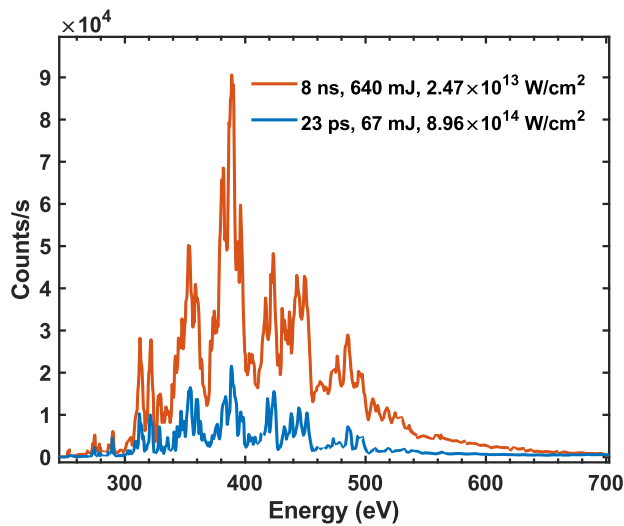

**Figure 2.** Comparison of Ar emission spectra for the ns and ps lasers, accumulated over 1 s integration time (20 pulses for the ps laser and 10 pulses for the ns laser).

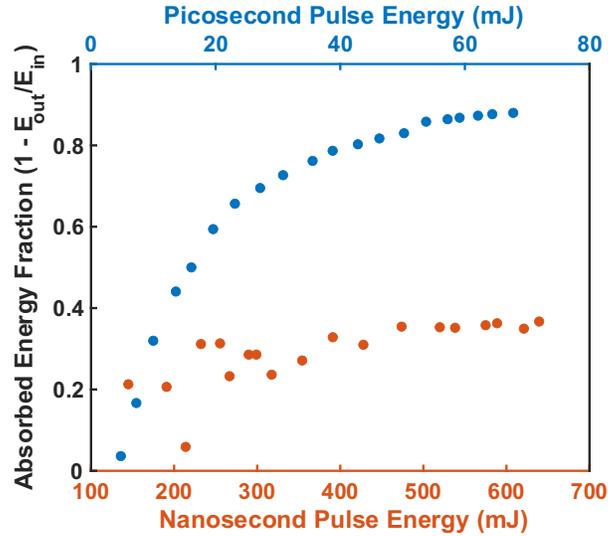

**Figure 3.** Coupling efficiencies for the ns and ps sources at varying input energies.

The coupling efficiency, defined as the fraction of the laser energy absorbed by the plasma, is compared for both the ps and ns sources in **Figure 3**. The ps source is coupled into the gaseous target at higher efficiencies, even at much lower input energies. This improved coupling efficiency has been observed in previous studies,[16] which reported similar trends when comparing plasmas in gaseous targets generated with 7 ns and 170 ps lasers.

Additionally, earlier work documented an increase in overall flux and a blue shift in the spectra from plasmas generated with ps lasers when the absorbed energy exceeded 200 mJ.[16] In this study, there does not appear to be a spectral shift towards higher energies with the ps plasma as it absorbs less energy. Comparison of the ps and ns sources here to those reported previously show a similar flux for our 23 ps source compared to the much higher pulse energy 170 ps source (**Figure 4**). This suggests that the over 8-fold increase in power density is compensated by the ~1/4 reduction in absorbed energy of the 23 ps source here compared to the 170 ps source in the resulting X-ray emission spectrum. It should be noted that the authors in this previous work used a 100 μm slit and 200 nm Al filter[16] (compared to the 50 μm slit and 100 nm Al filter used here).

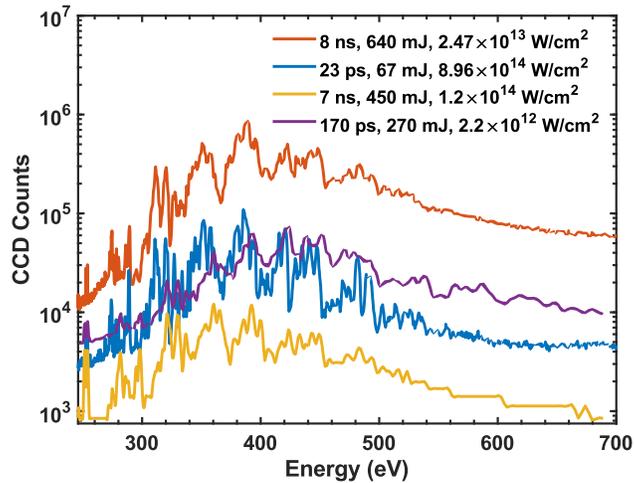

**Figure 4.** Comparison of the 8 ns and 23 ps sources presented in this study (red and blue, respectively) and the 170 ps and 7 ns sources presented in previous work (purple and yellow, respectively).[15] Data is presented as a sum over 100 pulses.

Alternatively, the increased energy density at the focal point of the ps laser could be resulting in a more strongly ionized plasma, but reabsorption by the surrounding gas plume could attenuate the signal that makes it to the detector. This is supported by the observation that the X-ray emission from the ps plasma drops off sharply when the pulse duration of the gas jet is increased beyond 550 μs. This suggests a strong degree of reabsorption of emitted X-rays by the surrounding gas plume. Previous reports have shown that ps sources produce smaller plasmas,[16,31] suggesting that the smaller size of the plasma could also play a role in the reabsorption of X-rays. A tungsten nozzle with a smaller diameter orifice that does not rapidly degrade in the plasma could address this issue and increase flux. Previous work on the gas delivery system has reported higher stability and increased X-ray flux.[24,32–34]

**Spectrometer Resolution and Sensitivity**

To quantify the spectral resolution of the instrument, the $N_2$ emission spectrum was collected with a 50 μm slit. The $N^{5+}$ emission peak at λ = 2.878 nm from the $N_2$ plasma emission spectrum was fit to a Gaussian with a FWHM of 1.02 eV, which gives a resolving power of E/ΔE = 424. The results of this fitting and the $N_2$ emission spectrum from the ps plasma integrated for 1 s are given in **Figure 5**.

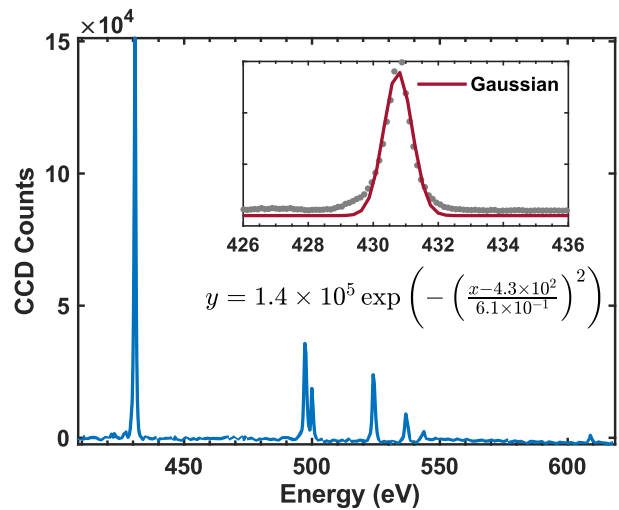

**Figure 5.** $N_2$ emission spectrum from ps SXR source and gaussian fit to the $N^{5+}$ emission peak at λ = 2.878 nm.

LPP sources have been shown to have high pulse-to-pulse instabilities and non-uniformity.[13,14] In the sources developed in this work, long-term degradation of the gas jet nozzle is observed due to the plasma generation process. In **Figure 6**, this degradation can be seen directly in loss of flux within the first few minutes of measurement. Over the course of 15 minutes (18,000 SXR pulses) intensity drops by 2.41% with a standard deviation over the mean of 1.02%.

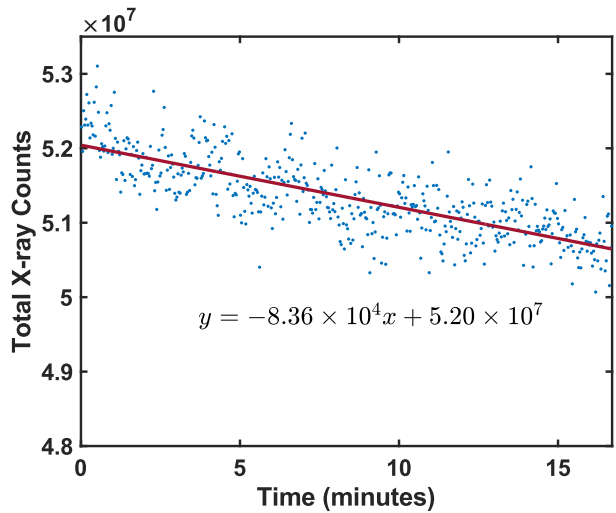

**Figure 6.** Total integrated X-ray intensity where each data point is 40 summed SXR shots as a function of time, fit to a linear regression.

**NEXAFS Spectra**

To quantify the system's key specifications for measuring X-ray absorption data, the absorption spectrum was measured for a commercially available 50 nm thick $Si_3N_4$ membrane and an ALD deposited 40 nm $TiO_2$ thin film. For the NEXAFS spectrum of $Si_3N_4$ and $TiO_2$, a 100 μm slit is used. **Figure 7** is the absorption spectra from the $TiO_2$ thin film at the Ti $L_{2,3}$ edges and the O K edge. **Figure 8** is the absorption spectrum of the $Si_3N_4$ membrane at the N K-edge. All spectra were accumulated over 3,000 SXR pulses, to give a total measurement time of 2.5 minutes, similar to or shorter than total acquisition times reported by similar LPP sources.[27,35,36] Absorption data were smoothed using locally weighted scatterplot (lowess) smoothing using a span of size 0.015. As can be seen by the Ti $L_{2,3}$ edges, the instrument's resolution can distinguish two transitions at 462 eV ($L_2$) and 465 eV ($L_3$), which exist due to spin-orbit coupling. The $Si_3N_4$ and $TiO_2$ absorption spectra show agreement with previous LPP sources[27] and synchrotron data,[37] respectively. The lack of splitting of the $L_2$ and $L_3$ edges in the measurements conducted in this work is observed because the TiO$_2$ sample prepared here is amorphous and was not annealed. The reference data is also amorphous TiO$_2$ measured at a synchrotron.[37]

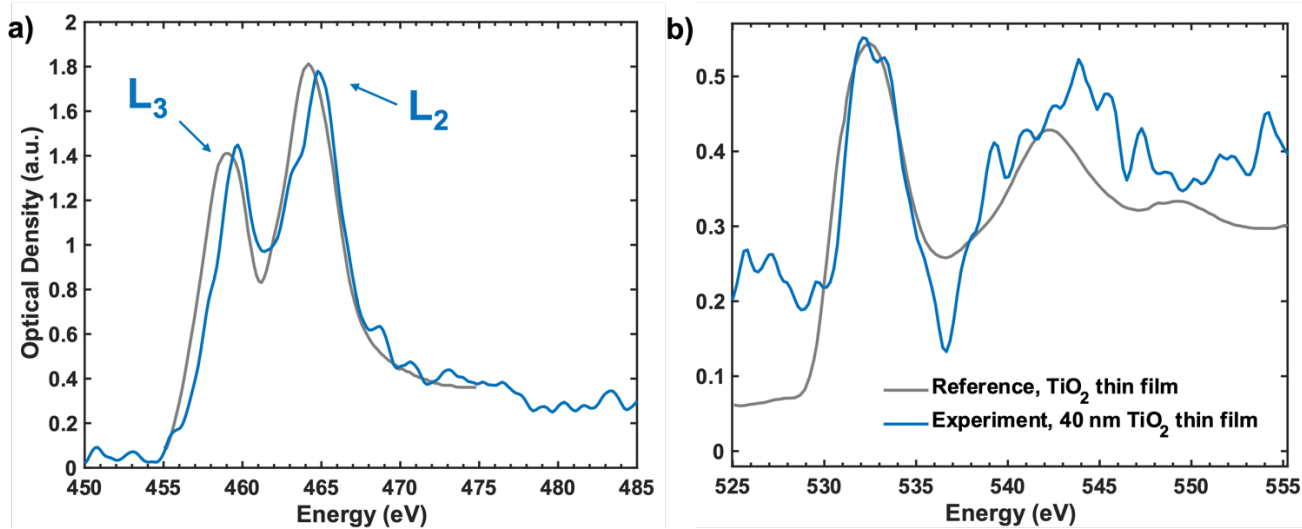

**Figure 7.** Absorption spectra of a 40 nm $TiO_2$ thin film at the **a)** Ti $L_{2,3}$ edges and **b)** O K edge. Reference spectra from ref [36] are synchrotron measurements of anatase $TiO_2$ and were scaled along the y-axis for comparison.

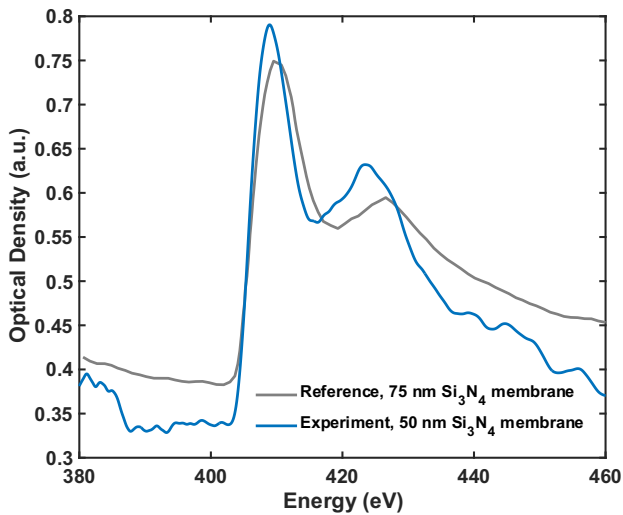

**Figure 8.** Absorption spectrum of the 50 nm $Si_3N_4$ membrane at the N K-edge compared to that of a 75 nm $Si_3N_4$ membrane from a LPP source in ref [26]. Reference data was not scaled along the y-axis.

## Conclusion

The key characteristics of X-ray emission from a laser-produced plasma source with a 23 ps driving laser were explored and benchmarked against an 8 ns system. A comparison with a much higher pulse energy ns source and similar sources in previously published reports shows the effect of shorter pulse duration and pulse energy on resulting X-ray spectra. The shorter pulses and resulting order of magnitude higher energy density at the target are not found to compensate for the lower total absorbed energy by the plasma in the resulting X-ray spectrum. However, the higher repetition rate still allows for the collection of similar NEXAFS spectra to other systems over similar measurement times. Furthermore, an X-ray grating with 3600 l/mm line density allowed for E/ΔE = 424 resolution. Although the stability of the X-ray emission is not high enough to collect long-term averages for time-resolved measurements due to gas jet nozzle degradation, improvements to the gas delivery system can be made to improve long-term stability.[24,32,33,38] Together, these capabilities make the LPP source a promising candidate for more complex X-ray absorption techniques, such as time-resolved and *in-situ* measurements.

## Acknowledgements

This work was funded by the Liquid Sunlight Alliance, which is supported by the U.S. Department of Energy, Office of Science, Office of Basic Energy Sciences, Fuels from Sunlight Hub under Award Number DE-SC0021266. This work was also funded by the Resnick Sustainability Institute and the Beckman Institute Pilot Program.

**Keywords:** X-ray absorption spectroscopy • Time-resolved spectroscopy • Analytical methods • Laser-produced plasmas • Table-top X-ray sources